# The creation of high-dimensional oscillations from low-dimensional systems

A.A.KIPCHATOV and L.V.KRASICHKOV

College of Applied Science, Saratov State University,

83 Astrakhanskaya, Saratov 410071, Russia

**Abstract** - The complication of chaotic oscillation under its transformation by linear inertial process is discussed. It is shown that such complication is begun from large scales of attractor and is pure dynamical process.

By now the main universalities of transition to chaos in low-dimensional nonlinear dynamical systems of different nature (maps, ordinary differential equations) have been well studied. Behaviour of such systems is characterized by low-dimensional attractors which are sufficiently simple in structure. For investigation of more high-dimensional chaotic oscillations, as a rule, it is considered either extended or coupled systems. However, there is another way of appearance of high-dimensional chaotic oscillations which has not been adequately studied. This way is a linear inertial transformation, or in other words, filtration of low-dimensional chaotic oscillations.

It is known that the linear filtration of chaotic oscillations leads to increase of dimension of output oscillation [1–3] and to complication of the fractal attractor structure [4]. This increase of complexity (i.e. dimension of oscillations) is the *deterministic dynamical process* and it is not connected with any noise.

To illustrate the deterministic nature of such complication we consider the linear inertial transformation on the example of simplest system, namely, the linear sum $s(t)$ of the chaotic oscillation $\xi(t)$ and the same oscillation taken with the time delay $\tau_c$:

$$s(t) = \xi(t) + \xi(t - \tau_c). \tag{1}$$

This process is similar to the band-stop filtration with transfer characteristic $H(\omega) = 1 + \exp(-j\pi\omega/\omega_0)$, where $\omega_0 = \pi\tau_c^{-1}$ is the frequency of maximum rejection. Generally speaking, $H(\omega)$ is periodic and has an infinite number of such frequencies $\omega_n = (2n+1)\omega_0, n = 0, 1, ....$

As chaotic signal $\xi(t)$ we used the oscillation of $x(t)$ variable of Rössler equations [5]

$$\dot{x} = -(y+z), \dot{y} = x + 0.2y, \dot{z} = 0.2 - 4.6z + xz. \tag{2}$$

Eq. (2) are integrated with the time step $dt = 0.04$ and we show, for further comparisons, the characteristics estimated from $x(t)$ in fig.1. Note that for the presented parameter values the system described by eq. (2) demonstrates low-dimensional chaos with sharply defined fundamental frequency and with very slow decay of correlation (see fig.2).

Throughout this paper, we estimate the phase portrait and the correlation dimension for the attractors reconstructed, from $x(t)$ or $s(t)$, on the basis of the time delay method [6]. The correlation dimension [7] is shown as the function of the length scale $E$, where $E = 20 \cdot \log_{10}(\varepsilon)$. The correlation integral is defined as in refs. [3,4,8,9] (i.e. $N$ is the number of data points and $M$ is the number of reference points). It is expected that $D_c(E) < 6$ and hence the selected value of embedding dimension, $d_E = 6$, is sufficient for estimations (see ref. [10]).

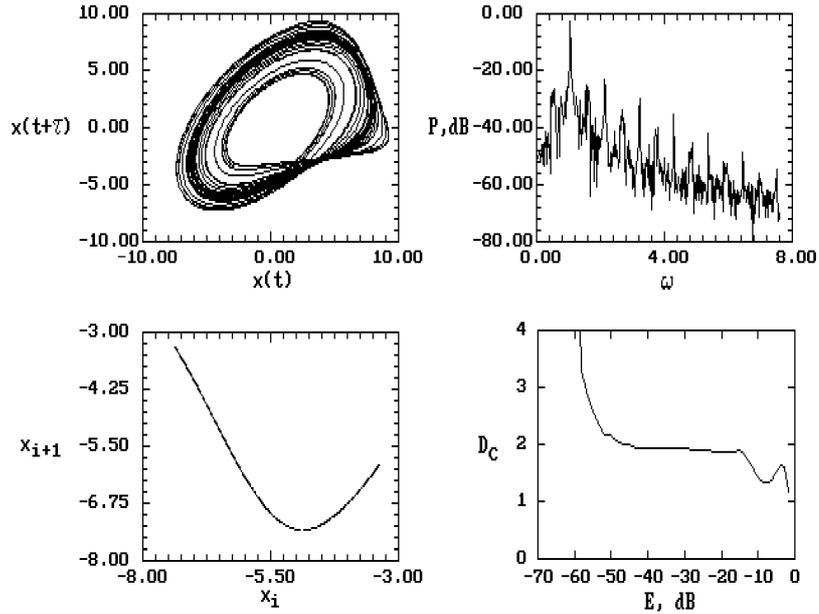

Fig. 1. The phase portrait (a), the power spectrum (b), the first return map (c) and the correlation dimension $D_c(E)$ (d) were estimated for $x(t)$ of the Rössler system eqs.(2). Parameters of the time delay and the correlation dimension methods are the following: $\tau = 1.0$, $d_E = 6$, $N = 10^5$, $M = 10^3$.

Consider the transformation of the signal $x(t)$ (2) by the process difined by eq. (1) for varying $\tau_c$ ($\xi(t) = x(t)$). Notice that $\tau_c$ will be selected in the minimums of the autocorrelation function (ACF) (fig.2). One might be see that, for such selection, one of frequencies of maximum rejection is equal to the fundamental fraquency ($\omega_s \approx 1.08$) of oscillations of the system eq. (2).

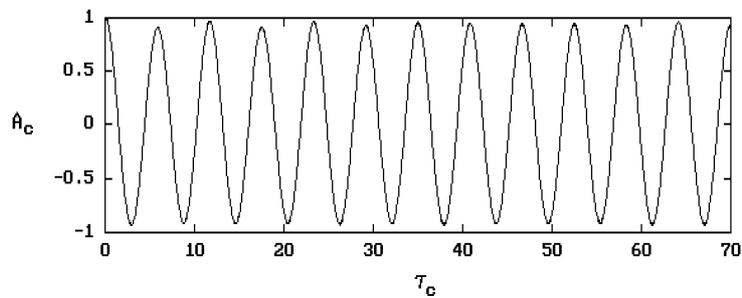

Fig. 2. The autocorrelation function for $x(t)$ of the Rössler system eqs.(2).

Fig.3 shows the characteristics of otput oscillation of process (1) for fifth ACF min, in this case, there are two frequencies of max rejection in $[0, \omega_s]$ for oscillation of the system (2). One can see from the first return map (fig.3c) that the chaotic oscillations are complicated due to stretching and folding of the attractor. The correlation dimension is increased for large length scales $-30\text{dB} \leq E \leq -10\text{dB}$ (fig.3d) and is unchanged for small ones. It exibits that the inner (thin) attractor structure is held.

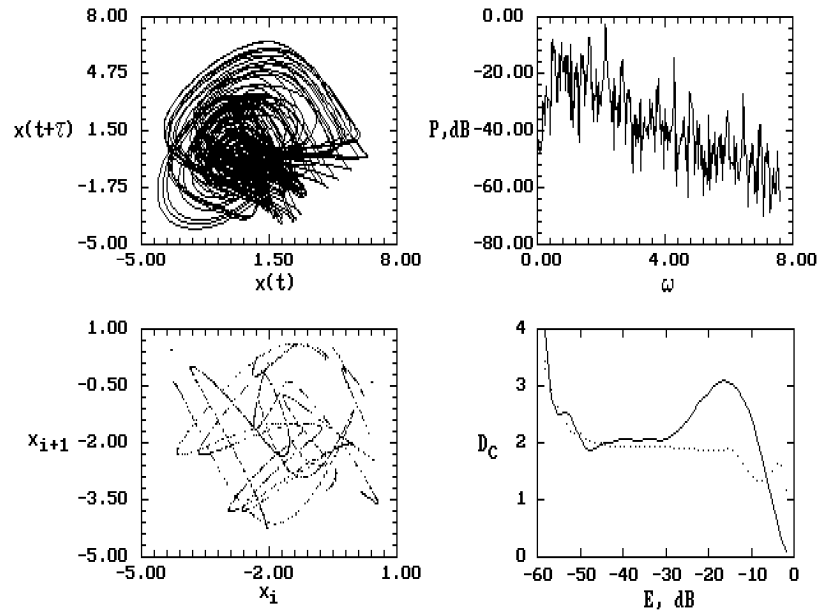

Fig. 3. Same characteristics as in Fig.1, but for $s(t)$ eq. (1) ($\tau_c = 654 dt = 26.16$). Dotted curve shows $D_c(E)$ from Fig.1d.

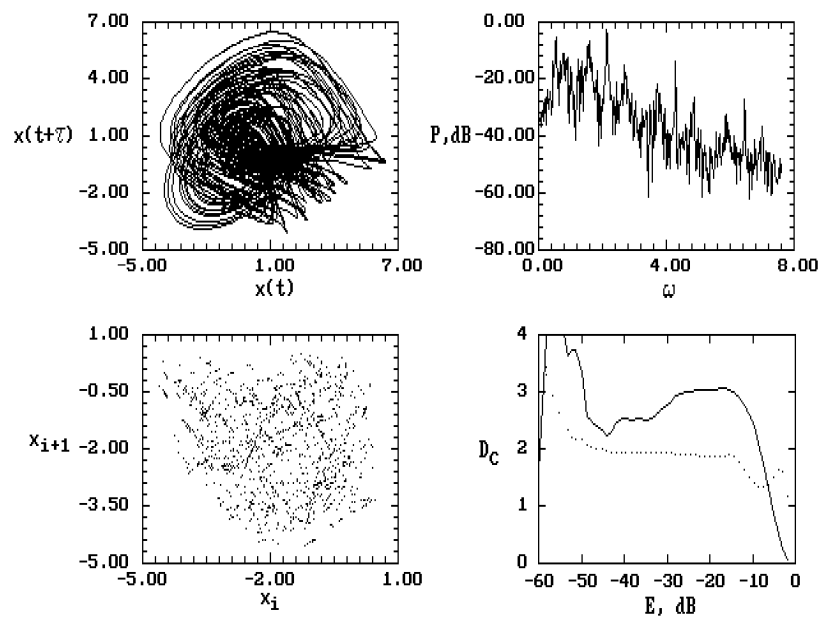

Fig. 4. Same characteristics as in Fig.3, but for $\tau_c = 1385 dt = 55.40$.

Strengthening of the inertial properties of the system, i.e. decreasing the correlations between the adding oscillations ($\tau_c$ in tenth ACF min and four frequencies of max rejection in $[0, \omega_s]$), leads to the many times folding of the attractor tapes and to arising the complicated attractor structure (fig.4c). The correlation dimension is increased for medium length scales (fig.4d). The further increasing of the time delay $\tau_c$ in eq. (1) leads to full mixing of points at the first return map and to significant increasing of dimension for wide range $E$, in which the dimension can be estimated reliably.

We emphasize that such complication of attractor, in presented case, is not connected with any noise. If noises were the main effects in the process, then $D_c(E)$ must be increased for small $E$. Whereas the obtained results show clearly the attractor complication from large to small $E$. Presented results allow to suppose that the inner attractor structure, even for very large values $\tau_c$, is not changed and the dimension of attractor for $E \to -\infty$ must be equal to that for original attractor of the input signal $\xi(t)$, i.e. behaviour of $s(t)$ is complitely determined by dynamics of eqs. (1) and (2). And in limit case, one might expect that for $\tau_c \to \infty$, or in other words, for adding (summing) of uncorrelated oscillations the dimension of sum of oscillations will be equal to the sum of dimensions of summands [11,12], and, for (1), this must lead to twice dimension.

It is necessary to stress that complication of chaotic oscillations under inertial transformation due to stretching and folding of the attractor types leading to the long scale changes of attractor is not unique way of the oscillation complication. In particular, the band-pass filtration leads to complication of attractor beginning from small scales ($E \to -\infty$) [4], and is connected with additional fractalization of attractor (fractal splitting of original attractor type).

All of discussed here phenomena have not still been investigated enough and invites further explainations. However, even our example shows that low-dimensional chaotic oscillations become more high-dimensional under its passege through the *linear* inertial systems. And we suppose that such process is one of possible *dynamical* ways to high-dimensional chaos in many real systems.

*Acknowledgments* – This work was supported by The Russian Fund of Fundamental Research under Grant No. 93-02-16171.